\documentclass[%
reprint,%
amssymb, amsmath,%
prm,cha,%
]{revtex4-1}

\usepackage[colorlinks=true,linkcolor=blue]{hyperref}%
\expandafter\ifx\csname package@font\endcsname\relax\else
\expandafter\expandafter
\expandafter\usepackage
\expandafter\expandafter
\expandafter{\csname package@font\endcsname}%
\fi
\usepackage[justification=justified]{caption}
\usepackage{multirow}
\usepackage{etoolbox}

\newcommand{\angstrom}{\mbox{\normalfont\AA}}
\BeforeBeginEnvironment{figure}{\vskip 0ex}
\AfterEndEnvironment{figure}{\vskip 0ex}

\usepackage[justification=justified]{subcaption}
\usepackage{graphicx}

\begin{document}
	
	\title{Oxygen-vacancy tuning of magnetism in  SrTi$_{0.75}$Fe$_{0.125}$Co$_{0.125}$O$_{3-\delta}$ perovskite}
	
	\author{M. A. Opazo$^1$, S. P. Ong$^2$, P. Vargas$^1$, C. A. Ross$^3$, and J. M. Florez$^{1,3}$}%
	\email{juanmanuel.florez@usm.cl}
	\affiliation{$^1$Departamento de F\'isica, Universidad T\'ecnica Federico Santa Mar\'ia, Casilla 110-V, Valpara\'iso, Chile\\
		$^2$Department of NanoEngineering, University of California, San Diego,
		9500 Gilman Drive, La Jolla, California 92093, USA\\
		$^3$Department of Materials Science and Engineering, Massachusetts Institute of Technology, 77 Massachusetts Avenue, Cambridge, Massachusetts 02139, USA}%
	
	%\date{March 2017}%
	%\revised{April 2017}%
	
	\begin{abstract}
	We use density functional theory to calculate the structure, band-gap and magnetic properties of oxygen-deficient SrTi$_{1-x-y}$Fe$_{x}$Co$_{y}$O$_{3-\delta}$ with $x$ = $y$ = 0.125 and $\delta = \{0, 0.125, 0.25\}$. The valence and the high or low spin-states of the Co and Fe ions, as well as the lattice distortion and the band-gap, depend on the oxygen deficiency, the locations of the vacancies, and on the direction of the Fe-Co axis. A charge redistribution that resembles a self-regulatory response lies behind the valence spin-state changes. Ferromagnetism dominates, and both the magnetization and the band gap are greatest at  $\delta$ = 0.125. This qualitatively mimics the previously reported magnetization measured for SrTiFeO$_{3-\delta}$, which was maximum at an intermediate deposition pressure of oxygen. \\
	
	\noindent
	Reprinted with permission from M. A. Opazo, S. P. Ong, P. Vargas, C. A. Ross and J. M. Florez.  	\href{https://journals.aps.org/prmaterials/abstract/10.1103/PhysRevMaterials.3.014404}{Physical Review Materials, 3, 014404 (2019)}.
	Copyright (2019) by the American Physical Society.

	\end{abstract}
	
	\maketitle

%%%%%%%%%%%%%%%%%%%%%%%%%%%%%%%%%%%%%%%%%%%%%%%%%%%%%%%%%%%%%%%%%%%%%%%%%%%%%%%%%%%%%%%%%%%%%%%%%%%%%%%%%%%%
%%%%%%%%%%%%%%%%%%%%%%%%%%%%%%%%%%%%%%%%%%%%%%%%%%%%%%%%%%%%%%%%%%%%%%%%%%%%%%%%%%%%%%%%%%%%%%%%%%%%%%%%%%%%
\section{Introduction}
Multiferroic materials provide a basis for a range of new devices proposed for information storage and processing, as well as offering a fertile ground for fundamental physics and materials science discoveries. In a multiferroic, two ferroic orders coexist, such as ferromagnetism and ferroelectricity, and coupling between them offers the opportunity for electrical control of magnetism, ideally at and above room temperature for device applications\cite{multiferroico1}. ABO$_3$ perovskite materials show a range of interesting electronic and ferroic phenomena including  superconductivity \cite{superiso_sto,supersemi_sto,superinter_bet_oxi,nonstoi_grain_sto}, strain-induced ferroelectricity in thin films and heterostructures \cite{Ferrooxide_superla,BiFeO_multi_hetero,BaTiOFerroel_film,Ferroelec_strin_sto}, isotope-exchange and electron-doping as well as point-defect and oxygen-vacancy-induced magnetism and ferroelectricity \cite{OdefSTF,Ferroindu_isoto_exchan_sto,anti-ferrodist_vacan_sto,oxy_vac_stco_teo-exp,magne_surfac_sto_charging,pointdefect_ferroelec_sto,large_magne_Odefici_sto,electro_doping_metal_oxide}, and room temperature multiferroicity  achieved through phase coexistence or oxygen deficiency stabilization \cite{OdefSTF,multiferro_roomt_ortho-morpho_lfo}.\\

SrTiO$_{3}$ (STO) is a prototypical perovskite, a paraelectric, non-magnetic large-band-gap semiconductor\cite{sto_with_hse} which has been incorporated into heterostructures that exhibit conductive interfaces or mixed bulk/surface order parameters \cite{OdefSTF,sto_with_hse,multiferroico1,Sikam:2018daa,Pai:2018fia}. There has been great interest in the role of both oxygen defects and substituents for Ti in triggering magnetism or ferroelectricity. STO can become ferroelectric under the application of uniaxial stress or static electric fields, as well as in the presence of defects or substituents that can disrupt the stability of the paraelectric phase by inducing charge fluctuations and rigid antiphase rotations of the oxygen octahedra. In the case of magnetism, replacing non-magnetic Ti$^{4+}$ ions with non-dilute magnetic transition metal  ions produces  magnetic behavior. The large band gap of STO suggests that defects will tend to introduce charges rather than shifting the Fermi energy to favor electron-gas-like states. \cite{OdefSTF, Mag-STO-delta, OrbitalSymmetry-STO, StrainControl, anti-ferrodist_vacan_sto, oxy_vac_stco_teo-exp, magne_surfac_sto_charging, pointdefect_ferroelec_sto, large_magne_Odefici_sto, electro_doping_metal_oxide, Sikam:2018daa, Dong:2018dr, Lee:2018dka, Pai:2018fia, Schiaffino:2017cwa, Brovko:2017hh, Wang:2017ii, added1, added2}. In STO, a general mechanism that explains the appearance of magnetic and ferroelectric properties and possible magnetoelectric behavior as a function of the presence of oxygen defects and B-site substituents is still lacking.\\

Here we focus on a theoretical study of oxygen-deficient STO in which Fe and Co are substituted for Ti to yield SrTi$_{1-x-y}$Fe$_{x}$Co$_{y}$O$_{3-\delta}$ (STFC). Recent experimental and theoretical results for SrTi$_{1-x}$Fe$_{x}$O$_{3-\delta}$ (STF) and SrTi$_{1-y}$Co$_{y}$O$_{3-\delta}$ (STC) show that their magnetic properties depend on both cation composition and O deficiency. For STC and STF, the magnetic moments increase with the Co or Fe content and with the O deficiency, though for STF the moment decreased again at the lowest growth pressures \cite{STC, OdefSTF,STFexp, roomt1, roomt2,FM-STC-delta}. In STF,  the substituents play an important role in determining the ferroelectric properties, which depend also on the band gap \cite{Brovko:2017hh}, while in STC, the band gap seems to be larger for the same concentration of Co. In STF, at very low oxygen deficiency, the FM and AFM orderings are close in energy with a mixture of high valence spin states, but in the STC case, a FM mixture of low and high spin states is predominant at similar oxygen deficiencies. The properties of STF and STC are very different in several aspects. The magnetization of STF depends on oxygen deficiency, but in STC, states in the band gap as a consequence of strong fractional-like hybridized occupations are less common, with more defined high occupations below the fermi energy \cite{STC,OdefSTF}. The multiferroic properties of O-deficient STO substituted with both Fe and Co are yet to be addressed. First principles modeling can give an understanding of the experimentally observed trends in the magnetic or ferroelectric observables as a function of the oxygen deficiency in many complex oxides, guiding the design of artificial multiferroic materials. These designs may be realized experimentally by manipulating the composition, microstructure, and strain state. \cite{multiferroico1,OdefSTF,STFexp, roomt1,roomt2,FM-STC-delta,Pai:2018fia}\\

In the present study, we  examine the stable valence states, charge distribution, band gap, structure, chemical strain and saturation magnetization of a material that combines the characteristics of STF and STC, with the aim of discovering  an O-deficiency-mediated STO-based multiferroic. We use first principles calculations to examine the effects of O deficiency and composition, $(\delta,x,y)$, on the electronic properties of STFC. This compound has not previously been studied using first-principles methods and the ranges of $(\delta,x,y)$ for which the material is a semiconductor with a magnetic or ferroelectric response are still to be identified. We use compositions that are similar to those used in recent experimental work on STF and STC.\\

This paper is organized as follows: in section 2 we present the model and methods. Section 3 describes the ground-state properties of $\delta=0$ systems without O vacancies. In section 4 the results for $\delta=0.125$ are presented and in section 5 the results for $\delta=0.25$ are presented. In section 6 the electronic properties, represented by the band gap and charge distribution, the magnetic moment as well as the structural changes as a function of $\delta$, are discussed. Finally, section 7 gives the conclusions. 

%%%%%%%%%%%%%%%%%%%%%%%%%%%%%%%%%%%%%%%%%%%%%%%%%%%%%%%%%%%%%%%%%%%%%%%%%%%%%%%%%%%%%%%%%%%%%%%%%%%%%%%%%%%%
%%%%%%%%%%%%%%%%%%%%%%%%%%%%%%%%%%%%%%%%%%%%%%%%%%%%%%%%%%%%%%%%%%%%%%%%%%%%%%%%%%%%%%%%%%%%%%%%%%%%%%%%%%%%
\section{Model and Methods}
All density functional theory (DFT) calculations in this work were performed using the Vienna Ab initio Simulation Package 
(VASP 5.4)\cite{vasp96prb, vaspbackground}. The screened hybrid Heyd-Scuseria-Ernzerhof (HSE06) functional was used\cite{HSE061} as it has been shown in previous works by the authors to yield fair agreement with experiments for the electronic properties of STC and STF\cite{STC, OdefSTF,sto_with_hse}. 
%%%%%%%%%%%%%%%%%%%%%%%%%%%%%%%%%%%%%%%%%%%%%%%%%%%%%%%%%%%%%%%%%%%%%%%%%%%%%%%%%%%%%%%%%%%%%%%%%%%%%%%%%%%%
\begin{figure}[t]
	\includegraphics[width=1\linewidth]{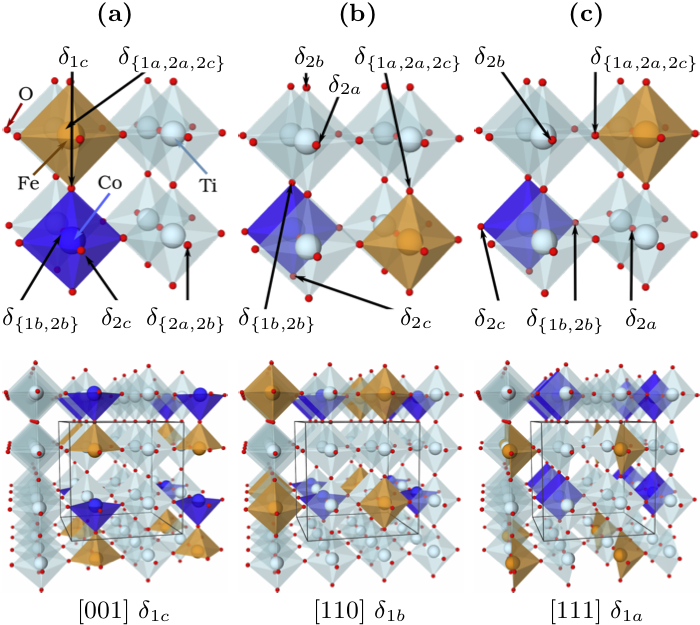}
	\captionof{figure}{Upper panels: cells of STFC with Fe-Co ion pairs located along: (a) [111]; (b) [110]; (c) [001].\\
		O-vacancies ($V_O$) are labeled \{$\delta_{1a}$, $\delta_{1b}$, $\delta_{1c}$\}, for three $V_O$ positions with one $V_O$/$u.c.$ corresponding to $\delta$ = 0.125. For two $V_O$/$u.c.$, where $\delta$ = 0.250, the three possible $V_O$-pair positions are labelled \{$\delta_{2a}$, $\delta_{2b}$, $\delta_{2c}$\}.\\ Lower panels: small ``single crystals'' having the respective $\delta_{1c}$, $\delta_{1b}$, $\delta_{1a}$ vacancies displayed in the upper panels. In all panels, (Fe, Co, Ti, O) = (brown, blue, light blue, red), and Sr is omitted for clarity.}
	\label{figure1}
\end{figure}
%%%%%%%%%%%%%%%%%%%%%%%%%%%%%%%%%%%%%%%%%%%%%%%%%%%%%%%%%%%%%%%%%%%%%%%%%%%%%%%%%%%%%%%%%%%%%%%%%%%%%%%%%%%%
%%%%%%%%%%%%%%%%%%%%%%%%%%%%%%%%%%%%%%%%%%%%%%%%%%%%%%%%%%%%%%%%%%%%%%%%%%%%%%%%%%%%%%%%%%%%%%%%%%%%%%%%%%%%
Spin-polarized calculations were performed using a plane-wave energy cutoff of 500 eV and a Monkhorst-Pack $2\times2\times2$ $k$-point grid on $2\times2\times2$ perovskite supercells (40 atoms for $\delta = 0$). One and two oxygen vacancies were introduced to obtain $\delta = 0.125$ and $0.25$ compositions, respectively, and  $x=y=0.125$. All symmetrically distinct vacancy/transition metal orderings within the supercell were calculated. Though larger supercells are useful in clarifying the role of lower symmetry arrangements, the high computational cost and the number of distinct structures make such calculations unreasonable to accomplish using the expensive HSE functional. Projector-augmented wave (PAW) PBE pseudopotentials were used with $3s^{2}4p^{6}5s^{2}$, $3d^{3}4s^{1}$, $3d^{7}4s^{1}$, $3d^{8}4s^{1}$, and $2s^{2}2p^{4}$ valence configurations for Sr, Ti, Fe, Co, and O respectively. All total forces were converged to within 0.05 eV \AA$^{-1}$. Most calculations were performed with neutral supercells for both stoichiometric and oxygen-deficient compositions. However, in order to test the main magnetic features of specific configurations, we have also considered charged one-vacancy supercells with a neutralizing background charge \cite{vaspbackground}.\\ 

Figure \ref{figure1} illustrates the multiple configurations that we analyzed in order to extract the STFC magnetic and electronic features. The upper panels show the possible $2\times2\times2$ arrangements of the Fe and Co ions  as well as the  vacancies. For our supercell, $\delta = 0.125$ requires the creation of one vacancy and $\delta = 0.25$ of two. The oxygens removed in each case are chosen at positions classified following symmetry rules with a 10${^{-4}}\text{\AA}$ tolerance. The lower panels in Figure \ref{figure1} illustrate how larger $4\times4\times4$ supercells would appear. The vacancies  are labelled as: $\delta_{{\#}{place}}$, with $\#=$ the number of vacancies, and place=\{a, b, c\} represents the type of vacancy in terms of its respective cation coordination. Among these coordination characteristics two are common to all the configurations i.e., we have incomplete octahedra with just one vacancy and at least one of the magnetic ions is located within an incomplete octahedron. These choices are associated with the maximization of the saturation magnetization/u.c. while reducing the possibility of  oxygen migration effects and strong interactions between vacancies. These criteria lead us to examine a subset of the statistical distribution of both the vacancies and the magnetic transition metals for our ABO$_{3-\delta}$ system. Finally, the structural, magnetic, and electronic analysis were performed using the Python Materials Genomics package \cite{pymatgen}, The Materials Project\cite{matproj}, VESTA \cite{VESTA}, OVITO \cite{ovito} and in-house packages.

%%%%%%%%%%%%%%%%%%%%%%%%%%%%%%%%%%%%%%%%%%%%%%%%%%%%%%%%%%%%%%%%%%%%%%%%%%%%%%%%%%%%%%%%%%%%%%%%%%%%%%%%%%%%
\section{Perovskite with $\delta=0$}
We start by analyzing the system with no vacancies, i.e. $\delta = 0$. For SrTi$_{0.75}$Fe$_{0.125}$Co$_{0.125}$O$_{3}$, the $2 \times 2 \times 2$ supercell contains one Co and one Fe ion which can be adjacent along [001], [110], or [111], separated by $a$, $\sqrt{2}a$ and $\sqrt{3}a$ respectively where a is the lattice parameter of STO, as illustrated in Figure \ref{figure1}.  Table \ref{table1} gives the structural, magnetic and band-gap properties of the three different arrangements of the Fe-Co axis. We observe from Table \ref{table1} that the [001] configuration is the ground state (gs) in STFC, being more than 200 meV lower in energy. This is unlike STC, where the [110] configuration is the ground state \cite{STC}.\\

%%%%%%%%%%%%%%%%%%%%%%%%%%%%%%%%%%%%%%%%%%%%%%%%%%%%%%%%%%%%%%%%%%%%%%%%%%%%%%%%%%%%%%%%%%%%%%%%%%%%%%%%%%%%
\begin{table}[b]
	\captionof{table}{Magnetic and structural properties  of SrTi$_{0.75}$Fe$_{0.125}$Co$_{0.125}$O$_{3}$.}
	\begin{tabular*}{\linewidth}{c@{\extracolsep{\fill}}cccccc}
		\hline \hline
		\multicolumn{2}{c}{Conf.}   &  $\mu_B$/Fe & $\mu_B$/Co & d$_{\text{\scalebox{0.85}{Fe-Co}}}$/d$_{\text{\scalebox{0.7}{STO}}}$ & $\Delta$E (meV) & E$_{bg}$(eV)\\ \hline
		\multicolumn{2}{l}{[111]}   &     3.76   &   1.04  &  0.988 & 230.8 & 0.332\\
		\multicolumn{2}{l}{[110]}&     3.74   &   1.04 & 0.984 &  221.7  & 0.185\\
		\multicolumn{2}{l}{[001]$_{gs}$} & 3.67 & 1.41 & 0.980 & - - - -  & 0.045 \\             \hline
		& &  a (\angstrom)& b (\angstrom) & a/c&  \multicolumn{2}{c}{$\mu_B$/u.c.} \\\hline
		&[001]$_{gs}$ & 7.751 & 7.752 & 0.994 & \multicolumn{2}{c}{5.03}\\
		\hline\hline
	\end{tabular*}
	\label{table1}
\end{table}
%%%%%%%%%%%%%%%%%%%%%%%%%%%%%%%%%%%%%%%%%%%%%%%%%%%%%%%%%%%%%%%%%%%%%%%%%%%%%%%%%%%%%%%%%%%%%%%%%%%%%%%%%%%%

\begin{figure}[t]
	\includegraphics[width=1\linewidth]{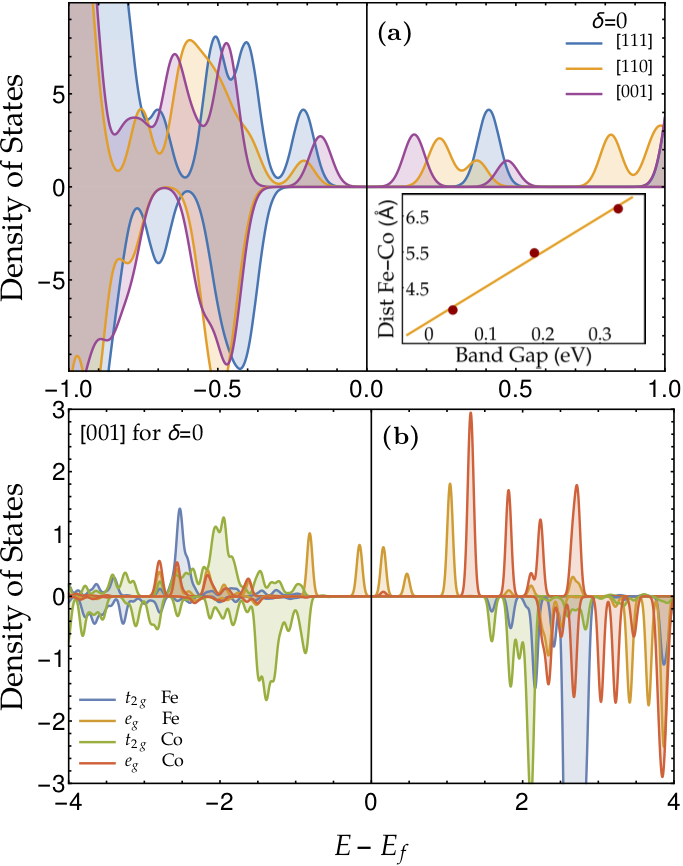}
	\caption{(a) DOS for states in Table \ref{table1}. Inset: Band-gap versus $d_{Fe-Co}$. (b) (Fe, Co) $t_{2g}e_g$ for [001]$_{gs}$ in Table \ref{table1}. Units of (eV) for Energy and (states/eV) for DOS.}
	\label{figure2}
\end{figure}

For $\delta = 0$, Co  is in a strongly hybridized $t_{2g}^{5}e_{g}^{0}$ 4+ low-spin state, and Fe is in what seems to be a $t_{2g}^{4}e_{g}^{2}$ 2+ high-spin state, similar to the ground states in $\delta=0$ STC\cite{STC} and STF\cite{OdefSTF,roomt1}. Fe\textsuperscript{4+} and Fe\textsuperscript{2+} high-spins have the same spin magnetization (4$\mu_B$) and Co\textsuperscript{4+} and Co\textsuperscript{2+} low-spin states both have 1$\mu_B$, therefore, if we judge the approximated molecular orbital states by the total magnetization, the systems should include one of four possible (Fe,Co) pairs. We see from Figure \ref{figure2} that the empty Co-$e_{g}$ and polarized Fe-$e_{g}$ requirements are characteristics of the states mentioned above. The two possible magnetic states for each ions differ by one electron per d-suborbital, hence, the slightly larger magnetic moment for Co (larger than 1$\mu_B$) is an indication of the O-mediated super-exchange coupling between Fe and Co, which end up sharing a spin of the 3d-Co, uncoupling a filled $t_{2g}$ that in turn is seen by the Fe ion through the oxygen.\\

The first intriguing result is that for $\delta=0$, STFC is a ferromagnetic (FM) semiconductor with a large magnetic moment of ~5 $\mu_B/$ per supercell and a band gap of $\approx50$ meV as shown in Table \ref{table1}. The inset of Figure \ref{figure2}(a) shows the increase of band gap $E_{bg}$ with the distance between Fe and Co. The transition metal ions, Fe and Co, mimic the electronic occupancy of the end-member perovskites, STF and STC, in the [111] configuration. If the Co and Fe ions were dispersed homogeneously into the STFC  and the O vacancy concentration (i.e. $\delta$) was very low,  these results predict that STFC will fulfill two key multiferroic requirements, i.e., to have a significant saturation magnetization and insulating behavior needed for ferroelectricity \cite{multiferroico1}. \\

%%%%%%%%%%%%%%%%%%%%%%%%%%%%%%%%%%%%%%%%%%%%%%%%%%%%%%%%%%%%%%%%%%%%%%%%%%%%%%%%%%%%%%%%%%%%%%%%%%%%%%%%%%%%
%%%%%%%%%%%%%%%%%%%%%%%%%%%%%%%%%%%%%%%%%%%%%%%%%%%%%%%%%%%%%%%%%%%%%%%%%%%%%%%%%%%%%%%%%%%%%%%%%%%%%%%%%%%%
The band-gap widening in the inset of Figure \ref{figure2}(a), as well as the slight increase of $\mu_B$/Fe versus $d_\text{Fe-Co}$, are dominated by the $d_{z^{2}}$ in the Fe-$e_{g}$ orbitals, which is similar to what happens to Ti in oxygen-deficient STO under pressure \cite{Mag-STO-delta}, see Figure \ref{figure2}(b); this orbital has a larger projection on the Fe-Co axis for [001] than for the other directions and therefore the lattice parameter is larger for that case, decreasing for [110] and [111]. This is in accordance with the Co magnetic moment, which is slightly larger for the [001] case due to the $p$-$d$ hybridization between Fe and Co when they are closer to each other. In-plane strain, engineered via lattice mismatch with the substrate, could be used to promote this interplay between $d_\text{Fe-Co}$ and Fe-$e_{g}$ by determining the out-of-plane lattice parameter and lattice distortion. Other effects, including the non-neutral charge state of the supercell will be discussed later.

%%%%%%%%%%%%%%%%%%%%%%%%%%%%%%%%%%%%%%%%%%%%%%%%%%%%%%%%%%%%%%%%%%%%%%%%%%%%%%%%%%%%%%%%%%%%%%%%%%%%%%%%%%%%
%%%%%%%%%%%%%%%%%%%%%%%%%%%%%%%%%%%%%%%%%%%%%%%%%%%%%%%%%%%%%%%%%%%%%%%%%%%%%%%%%%%%%%%%%%%%%%%%%%%%%%%%%%%%
\section{Oxygen deficient $\delta = 0.125$ systems}
Now we consider the effects of oxygen deficiency. Previous studies in STF and STC have shown that the local spin magnitude tends to increase when a vacancy is immediately adjacent to the Fe or Co substituents \cite{OdefSTF, STFexp, STC}. Here, starting from the perovskites with $\delta=0$  in Table \ref{table1}, we relax the structure after introducing the vacancies indicated in Figure \ref{figure1}. In general, FM states are favored though AFM states can compete at certain vacancy and Fe contents. For instance, for the states in Table \ref{table1}, the AFM states are higher in energy by 217, 229, and 432 meV/f.u. for the [001], [110], and [111] Fe-Co directions, respectively. Therefore, we focus here on FM solutions.\\

\begin{figure}[t]
	\includegraphics[width=1\linewidth]{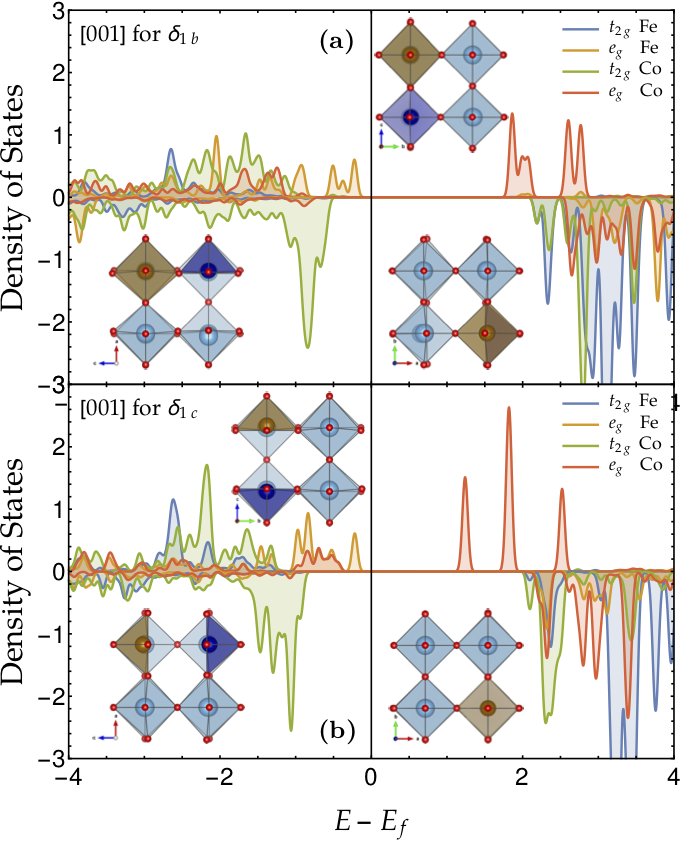}
	\caption{$t_{2g}$$e_g$ of Fe and Co for (a) [001]-$\delta_{1b}$ and (b) $\delta_{1c}$ in Table \ref{table2}. Insets: orthogonal views of the $2 \times 2 \times 2$ cells. Units of (eV) for Energy and (states/eV) for DOS.}
	\label{figure3}
\end{figure}

%%%%%%%%%%%%%%%%%%%%%%%%%%%%%%%%%%%%%%%%%%%%%%%%%%%%%%%%%%%%%%%%%%%%%%%%%%%%%%%%%%%%%%%%%%%%%%%%%%%%%%%%%%%
\begin{table}[b]
	\captionof{table}{Magnetic and structural properties of SrTi$_{0.75}$Fe$_{0.125}$Co$_{0.125}$O$_{2.875}$.}
	\begin{tabular*}{\linewidth}{c@{\extracolsep{\fill}}lcccrc}
		\hline \hline
		\multicolumn{2}{c}{Vacancy}   &  $\mu_B$/Fe & $\mu_B$/Co & $\mu_B$/u.c. &E(meV) & E$_{bg}$(eV)\\ \hline
		&[111]$\delta_{1a}$ & 4.12& 0.00 & 4.882 & 141.063 & 2.104 \\
		\multirow{8}{*}{} &[111]$\delta_{1b}$  &4.19 & 1.87 & 6.860  & 223.949 & 1.274\\
		
		&[110]$\delta_{1a}$ & 4.11&  0.00 & 4.880 & 97.075 & 2.112 \\
		&[110]$\delta_{1b}$ & 4.19& 1.85& 6.856& 129.492 & 1.312 \\
		$\quad$&[001]$\delta_{1a}$   & 4.08& 0.11& 4.869 & 326.112 & 1.991 \\
		&[001]$_{gs}$$\delta_{1b}$  & 4.18& 1.92&   6.840 & - - - - - &  1.811\\
		&[001]$\delta_{1c}$ & 4.11 & 1.85 &6.780     & 39.240 & 1.186
		\\             \hline
		& & a (\angstrom) & b (\angstrom)& a/c&  \multicolumn{2}{c}{d$_{\text{\scalebox{0.85}{Fe-Co}}}$/d$_{\text{\scalebox{0.7}{STO}}}$} \\\hline
		& [001]$_{gs}$($\delta_{1b}$)  &7.840 & 7.784 & 1.005 & \multicolumn{2}{c}{0.988}\\
		\hline\hline
	\end{tabular*}
	\label{table2}
\end{table}
%%%%%%%%%%%%%%%%%%%%%%%%%%%%%%%%%%%%%%%%%%%%%%%%%%%%%%%%%%%%%%%%%%%%%%%%%%%%%%%%%%%%%%%%%%%%%%%%%%%%%%%%%%%%
In Table \ref{table2} we present the results for selected cases of $\delta=0.125$, i.e. 1-V$_O$/$u.c.$. The vacancy locations are indicated in Figure \ref{figure1} as $\delta_{1a}$, $\delta_{1b}$ and $\delta_{1c}$ with ``1" meaning the number of vacancies and $a, b, c$ stand for: $a=$ vacancy between Fe and Ti, $b=$ vacancy between Co and Ti, and $c=$ vacancy between Fe and Co.
Two  cases are shown for a vacancy placed in the supercell with a Fe-Co axis of [111] and [110]; and three cases for a vacancy placed in the supercell with a Fe-Co axis of [001] as specified above. The additional [001] case corresponds to one  in which the vacancy is shared by Fe and Co. From Table \ref{table2} we see that in the O-deficient case, [001] remains as the ground state configuration, which is remarkably different from the rest of the O-deficient configurations because its lowest energy state has the vacancy located adjacent to the Co ion (vacancy $\delta_{1b}$) whereas in the [111] and [110] the lowest energy configuration has the vacancy adjacent to the Fe (vacancy $\delta_{1a}$). The [001] $gs$  shows an enhanced band-gap compared to $\delta=0$ as well as higher magnetic moments for both Fe and Co, maximizing the $\mu_B/u.c.$. This behavior is also observed in the second lowest energy state ([001] $\delta_{1c}$) which is higher than the $gs$ by $\approx40$ meV, with a band gap that is $30\%$ smaller as depicted in Figure \ref{figure3}.\\

%%%%%%%%%%%%%%%%%%%%%%%%%%%%%%%%%%%%%%%%%%%%%%%%%%%%%%%%%%%%%%%%%%%%%%%%%%%%%%%%%%%%%%%%%%%%%%%%%%%%%%%%%%%%
%%%%%%%%%%%%%%%%%%%%%%%%%%%%%%%%%%%%%%%%%%%%%%%%%%%%%%%%%%%%%%%%%%%%%%%%%%%%%%%%%%%%%%%%%%%%%%%%%%%%%%%%%%%%
The densities of states (DOS) for both the [001] $gs$ ($\delta_{1b}$) and next lowest energy state ($\delta_{1c}$)  are shown in Figure \ref{figure3}. These have incomplete coordination of the Co, which is in high spin Co$^{2+}$ with $t_{2g}^{5}e_{g}^{2}$, and Fe is in a high Fe$^{3+}$ with $t_{2g}^{3}e_{g}^{2}$ in the ground state. In the next lowest energy state, $\delta_{1c}$, Co is in a low spin Co$^{2+}$ with $t_{2g}^{6}e_{g}^{1}$ and Fe in a high spin Fe$^{3+}$ with $t_{2g}^{3}e_{g}^{2}$. This is consistent with the position of the $\delta_{1c}$ vacancy that is shared by Co and Fe, and the magnetic moments in Table \ref{table2} which slightly decrease from $gs$ to the next lowest energy mostly as a consequence of the Fe-Co $t_{2g}$ orbitals. In contrast the remaining two configurations, [110] and [111], favor an incomplete octahedral coordination of Fe (vacancy $\delta_{1a}$) but with a large energy cost.  The lowest energy state of both is a high spin Fe$^{3+}$ with $t_{2g}^{3}e_{g}^{2}$ and a low spin Co$^{3+}$ corresponding to $t_{2g}^{6}e_{g}^{0}$.\\

%%%%%%%%%%%%%%%%%%%%%%%%%%%%%%%%%%%%%%%%%%%%%%%%%%%%%%%%%%%%%%%%%%%%%%%%%%%%%%%%%%%%%%%%%%%%%%%%%%%%%%%%%%%%
\begin{figure}[t]
	\includegraphics[width=1\linewidth]{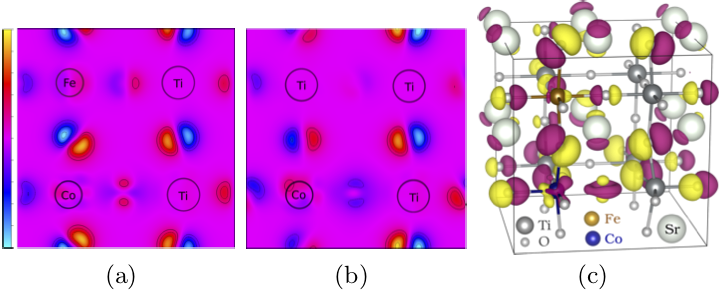}
	\caption{Charge density difference for [001] $\delta_{1b}$: CHG$_{[001]}$ - CHG$_{[001] \delta_{1b}}$ - CHG$_\text{Removed Oxygen}$. With CHG$_{X}$ being the charge density of system $X$. In (c), the 3D density-charge is displayed, with pink and yellow lobes of charge values +0.05 and -0.05 respectively. The charge density is projected in: (a) the lower horizontal  plane containing Co and V$_O$, and (b) the front vertical plane containing Fe, Co and V$_O$.}
	\label{figure4}
\end{figure}
%%%%%%%%%%%%%%%%%%%%%%%%%%%%%%%%%%%%%%%%%%%%%%%%%%%%%%%%%%%%%%%%%%%%%%%%%%%%%%%%%%%%%%%%%%%%%%%%%%%%%%%%%%%%
%%%%%%%%%%%%%%%%%%%%%%%%%%%%%%%%%%%%%%%%%%%%%%%%%%%%%%%%%%%%%%%%%%%%%%%%%%%%%%%%%%%%%%%%%%%%%%%%%%%%%%%%%%%%
\begin{figure}[b]
	\centering
	\includegraphics[width=\linewidth]{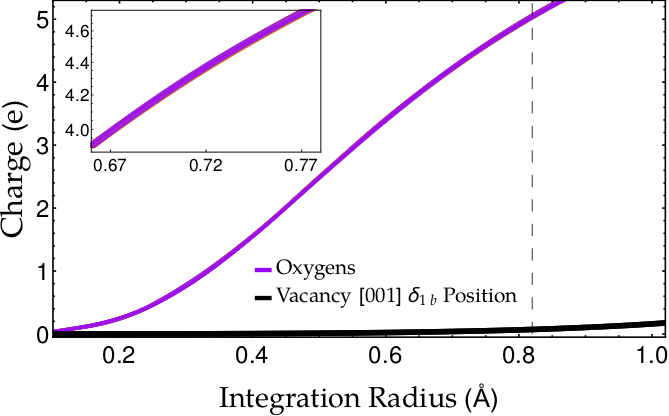}
	\caption{Total charge integrated at; the locations of the oxygens (all converge to the purple line) and at the center of the Vo (lower black line). The  Wigner-Seitz oxygen radius is displayed as a dashed vertical line.}
	\label{figure5}
\end{figure}
%%%%%%%%%%%%%%%%%%%%%%%%%%%%%%%%%%%%%%%%%%%%%%%%%%%%%%%%%%%%%%%%%%%%%%%%%%%%%%%%%%%%%%%%%%%%%%%%%%%%%%%%%%%%

To get a further insight on the resulting charge distribution behind the valence states, we display in Figure \ref{figure4} a charge-density difference that illustrates the charge redistribution that takes places when we go from a system with $\delta=0$ to a [001] $\delta_{1b}$ deficient perovskite. The volumetric plot in Figure \ref{figure4} (c), shows the resulting re-hybridization experienced by the metallic ions as a consequence of the necessity of rearranging the electrons that previously linked the now-missing oxygen. The projected charge densities displayed in (a) and (b) in the same figure show $d_{z^{2}}$-like $e_{g}$-hybridized-orbitals, which is an indication of the different $e_{g}$ populations between the $\delta=0$ perovskite and deficient systems, as the first case must have a $e_{g}^{0}$ and the second one has an unpaired electron in $d_{z^{2}}$; this difference is observed at the missing oxygen site through the p-d hybridization of the $\delta=0$ perovskite Co-O bonds. However, the vacancy itself is negligibly charged as  demonstrated by Figure \ref{figure5}, which compares the radially-integrated charge at the oxygen sites with the charge at the ``center" of the vacancy. The reorganization  of the available charge mimics the charge balance and self-regulation of transition metals in halide perovskites, rare-earth nickelates, embedded or semiconducting crystals and titanium-based oxides \cite{selfregu_changeoxida_insulator,chargedensi_vs_oxida_states, oxidation_partialcharge_ionicity,Varignon:2017is}, nevertheless, as we discuss below, the electronic-charge changes responsible for the resulting mixed valence spin states do not impede the self-regulation-like mechanism of the transition-metal total charge.\\
%%%%%%%%%%%%%%%%%%%%%%%%%%%%%%%%%%%%%%%%%%%%%%%%%%%%%%%%%%%%%%%%%%%%%%%%%%%%%%%%%%%%%%%%%%%%%%%%%%%%%%%%%%%%
\begin{figure}[t]
	\includegraphics[width=1\linewidth]{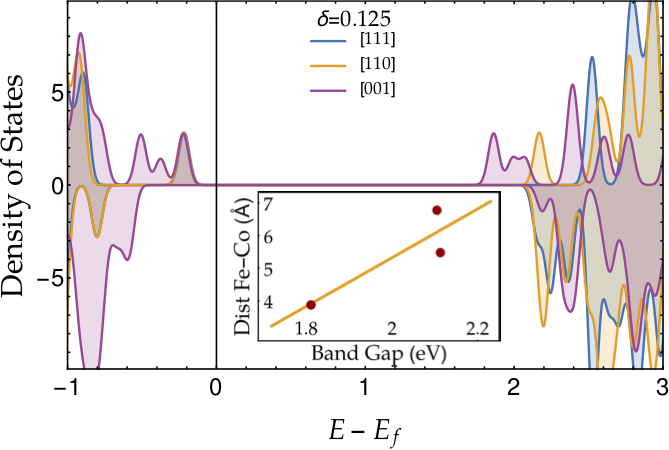}
	\caption{DOS for [001]$_{gs}$ $\delta_{1b}$, [110]$\delta_{1a}$ and [111]$\delta_{1a}$ in Table \ref{table2}.
		Inset: Band-gaps versus $d_{Fe-Co}$. Units of (eV) for Energy and (states/eV) for DOS.}
	\vspace{1pt}
	\label{figure6}
\end{figure}
%%%%%%%%%%%%%%%%%%%%%%%%%%%%%%%%%%%%%%%%%%%%%%%%%%%%%%%%%%%%%%%%%%%%%%%%%%%%%%%%%%%%%%%%%%%%%%%%%%%%%%%%%%%%
These results may be compared with the case of two Co and one V$_O$ in the supercell, $x=0$, $y=0.25$ (STC), where the two Co present two different valence spin states because one is adjacent to the vacancy and the other is not \cite{STC}. For two Fe and one V$_O$ in the supercell, $x=0.25$, $y=0$ (STF), there is also a mixed valence \cite{OdefSTF}, though for the Co case the FM configurations are the lower energy states and in the Fe case it is the AFM configurations. The inset in Figure \ref{figure6} shows that, as in the $\delta=0$ case, the band gap of STFC increases as $d_{Fe-Co}$ goes from the [001] to the [111] or [110] nearest neighbor configurations, but the difference between the latter two are small and are limited by the Fe-t$_{2g}$ orbitals at the valence band, and influenced by the shift of empty Co-e$_g$ levels. The aforementioned relationship between the Fe-Co axis and the tilted $d_{z^{2}}$ orbitals applies here just to the [111]-[110] comparison. The [001] gap is determined by the partially empty Co-$d_{x^{2}-y^{2}}$ orbital and the half-filled Fe-$e_{g}$, Figure \ref{figure3}. V$_O$ are located at Co-octahedra for [001] and at Fe-octahedra for [111] and [110], explaining the difference between the Co magnetic moments in Table \ref{table2}. The $gs$ band gap is large and comparable to the values predicted for STC ($x=0$, $y=0.25$) \cite{STC} and slightly smaller compared to recently reported band gaps for STF ($x=0.125$, $y=0$) with $\delta=0.125$ \cite{OdefSTF, STFexp}. These features suggest STFC as a possible magnetic semiconductor.\\

Our first-principles calculations in O-deficient STC (not shown here) predict the existence of electric polarization  in which the non-centrosymmetric redistribution of the charge was accompanied by complex lattice distortions beyond rigid $O_{6}$ rotations. Similar distortions are predicted for STFC as shown in the insets of Figure \ref{figure3}, and this suggests that STFC may be a promising system in which to explore multiferroicity.

%%%%%%%%%%%%%%%%%%%%%%%%%%%%%%%%%%%%%%%%%%%%%%%%%%%%%%%%%%%%%%%%%%%%%%%%%%%%%%%%%%%%%%%%%%%%%%%%%%%%%%%%%%%
\section{Oxygen deficient $\delta=0.25$ systems}
We next discuss the case of STFC with $\delta=0.25$, shown in Table \ref{table3}. Three different types of vacancy-pairs are studied for each Fe-Co axis orientation in Table \ref{table1}. Each corresponds to two O vacancies in the supercell, and the locations of three possible pairs of vacancies are shown in Figure \ref{figure1} as $\delta_{2a}, \delta_{2b}$ and $\delta_{2c}$, with the pair of vacancies between Fe-Ti and Ti-Ti, Co-Ti and Ti-Ti, and Fe-Ti and Co-Ti respectively.\\

%%%%%%%%%%%%%%%%%%%%%%%%%%%%%%%%%%%%%%%%%%%%%%%%%%%%%%%%%%%%%%%%%%%%%%%%%%%%%%%%%%%%%%%%%%%%%%%%%%%%%%%%%%%%
\begin{table}[b]
	\captionof{table}{Magnetic and structural properties for SrTi$_{0.75}$Fe$_{0.125}$Co$_{0.125}$O$_{2.75}$. }
	\begin{tabular*}{\linewidth}{c@{\extracolsep{\fill}}lcccrc}
		\hline \hline
		\multicolumn{2}{c}{Conf.}   &  $\mu_B$/Fe & $\mu_B$/Co &$\mu_B$/u.c. &E(meV) & E$_{bg}$(eV)\\ \hline
		
		&[111]$\delta_{2a}$& 3.58 & 0.92 &4.87 &188.95& 1.895 \\
		&[111]$\delta_{2b}$ & 4.19 & 0.90 &6.48 &645.89& 0.327\\
		
		\multirow{8}{*}{  }
		&[111]$\delta_{2c}$& 4.12 & 0.72 &5.26 &247.95& 0  \\
		&[110]$\delta_{2a}$ & 4.11 & 0.92 &5.24 &1126.26& 0.179 \\
		&[110]$\delta_{2b}$& 4.13 & 0.91 &5.65 &714.62& 0.129 \\
		&[110]$\delta_{2c}$& 4.11 & 0.74 &5.26 &249.49& 0  \\
		&[001]$\delta_{2a}$& 3.73 & 0.12 &3.46 &1445.42& 0.124 \\
		
		&[001]$\delta_{2b}$& 4.16 & 0.97 &5.18 &864.44& 0.351 \\
		&[001]$_{gs}$ $\delta_{2c}$& 3.78 & 0.95 &5.03 & - - - - - & 0.709 \\
		
		\hline
		& & a (\angstrom) & b (\angstrom) & a/c&  \multicolumn{2}{c}{d$_{\text{\scalebox{0.85}{Fe-Co}}}$/d$_{\text{\scalebox{0.7}{STO}}}$} \\\hline
		& [001]$_{gs}$($\delta_{2c}$)  &7.784 & 7.766 & 0.999 & \multicolumn{2}{c}{0.994}\\
		\hline\hline
		\label{table3}
	\end{tabular*}
	\vspace{-0.45cm}
\end{table}
%%%%%%%%%%%%%%%%%%%%%%%%%%%%%%%%%%%%%%%%%%%%%%%%%%%%%%%%%%%%%%%%%%%%%%%%%%%%%%%%%%%%%%%%%%%%%%%%%%%%%%%%%%%%

%%%%%%%%%%%%%%%%%%%%%%%%%%%%%%%%%%%%%%%%%%%%%%%%%%%%%%%%%%%%%%%%%%%%%%%%%%%%%%%%%%%%%%%%%%%%%%%%%%%%%%%%%%%%
\begin{figure}[t]
	\includegraphics[width=1\linewidth]{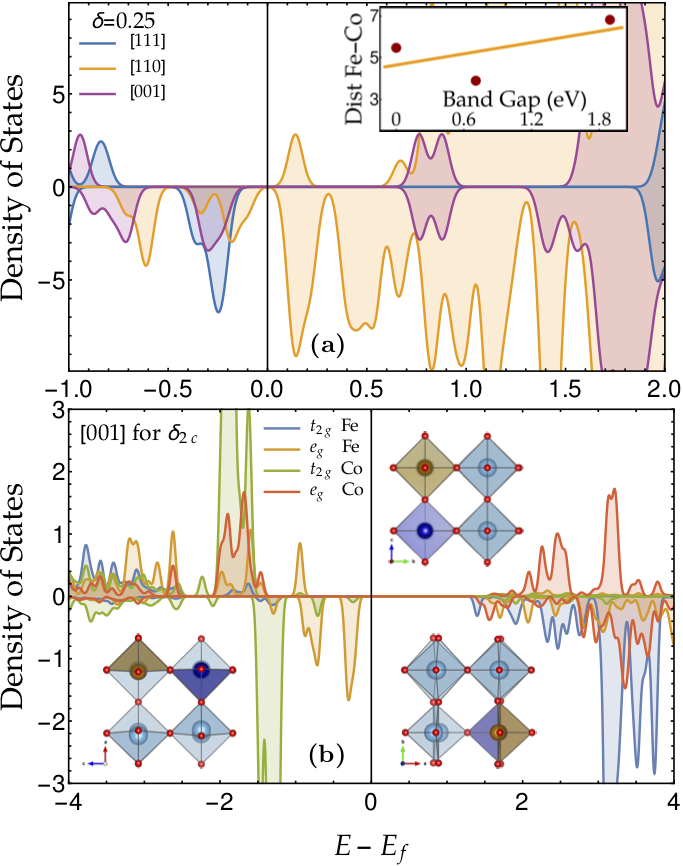}
	\caption{(a) DOS for ${gs}$-vacancies for arrangement in Table \ref{table3}.
		(b) (Fe, Co) $t_{2g}e_g$ for [001]$_{gs}$ $\delta_{2c}$ in Table \ref{table3}. Units of (eV) for Energy and (states/eV) for DOS.}
	\label{figure7}
\end{figure}
%%%%%%%%%%%%%%%%%%%%%%%%%%%%%%%%%%%%%%%%%%%%%%%%%%%%%%%%%%%%%%%%%%%%%%%%%%%%%%%%%%%%%%%%%%%%%%%%%%%%%%%%%%%%
The $gs$ corresponds to the system with the Fe-Co axis along [001] and the vacancies in  the  $\delta_{2c}$ sites. This $gs$ is indeed a semiconductor though it presents a narrower band-gap than the $\delta=0.125$ case but still wider than the $\delta=0$ case. The local magnetic moments of Fe and Co ions have decreased, as has the net magnetization $\mu_{B}/u.c.$. Intuitively this is a consequence of the vacancy location $\delta_{2c}$ for [001], i.e., both Fe and Co are coordinated by a vacancy. The next lowest energy state is the [111] case with vacancy locations $\delta_{2a}$,  i.e. one Fe and Ti are coordinated by a vacancy, which is also a semiconductor with a large band gap as depicted in Figure \ref{figure7}. In the inset (a) of this figure we can see the band-gap vs. Fe-Co spacing for the lowest energy states of $[001]$, $[110]$, and $[111]$ in Table \ref{table3}. The band gap decreases from the [001] to the [110] configuration for $\delta_{2c}$,  but it increases for the [111] configuration for $\delta_{2a}$. The lowest energy state for the [111] configuration is a semiconductor with a high spin Fe$^{2+}$ with $t_{2g}^{4}e_{g}^{2}$ occupation. For [110] the lowest energy state is metallic with Fe again dominating the conductivity,  with a $t_{2g}^{4}e_{g}^{2}$ distribution; Co$^{2+}$ seems to be closer to $t_{2g}^{5}e_{g}^{2}$ than to $t_{2g}^{6}e_{g}^{1}$ as expected.\\

The band-gap of the ${gs}$ is determined by a half-filled Fe-$e_{g}$ that reflects the high spin Fe$^{2+}$ $t_{2g}^{4}e_{g}^{2}$ state and could be inferred from Figure \ref{figure7}. On the other hand, the $gs$ has a low spin Co$^{2+}$ with $t_{2g}^{6}e_{g}^{1}$. This result for Co turns out to be contrary to what has been observed for STC and STF with one vacancy surrounding a $3d$ transition metal in terms of the resulting total magnetic moment. The Co charge self-organization in the presence of the vacancy suggests sharing between $t_{2g}$ and $e_{g}$, and such sharing contributes to the non-zero magnetization of the Co. \\

Such partial occupation of the $e_{g}$ orbitals is a feature consistently observed in the trends extracted out of systematic analysis of hundreds of crystal configurations, as are the charge states observed through our transition metal DFT studies. The saturation magnetization representing those hybridized features are often well defined by the local orbital-magnetism of the solid solutions \cite{oxidation_partialcharge_ionicity}. Our results on STFC do not display the fractional valences observed in TiO$_{2}$ \cite{oxidation_partialcharge_ionicity}, which might be a valid question for large vacancy concentrations in STFC with $x=y=$0.125, where we expect Ti-vacancy-Ti and O$_{4}$-coordinated Ti to be abundant. Instead, the Fe or Co ions respond to the oxygen vacancies, which both have experimentally been related to fractional valences \cite{jmflorezmagnetite,STC}.\\

\begin{figure}[t]
	\includegraphics[width=1\linewidth]{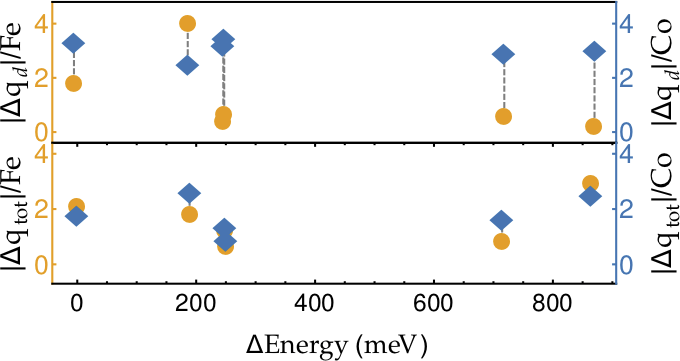}
	\caption{Total and d-orbital charge variation for the two lowest-energy configurations of each Fe-Co alignment in Table \ref{table3} measured as a percentage with respect to the $\delta=0$ charge, for the Fe (orange circles) and the Co (blue diamonds).}
	\label{figure8}
\end{figure}

The $d$-orbital hybridization processes that are adopted by the system to promote heteropolar bonding can be seen intuitively  from Figure \ref{figure8}, which presents the changes in the total and $d$-orbital charge of the Fe and Co ions in the two lowest-energy states for every Fe-Co distribution in Table \ref{table3}, for $\delta=0.25$. Figure \ref{figure8} shows the change in charge measured as percentage with respect to the respective $\delta=0$ state. 
We can see, independently of the stabilized distribution of vacancies, that there is a small  difference in the change of total charge between Fe and Co. This difference, besides reflecting the fact that the two vacancies are always coordinating at least one of the magnetic ions, and the relatively narrow dispersion of the total charges observed in the lower panel of Figure \ref{figure8} could be taken as fingerprints of a self-regulatory response  \cite{chargedensi_vs_oxida_states} of the transition metals in our O deficient perovskites. However, a self-regulatory interpretation does not conflict with the aforementioned STFC charge-redistribution process, which gives rise to the magnetic behavior, i.e., the magnetic order parameter is dominated by the 3d valence-electrons and as the upper panel in Figure \ref{figure8} suggests, the 3d-orbital charges do change in both Fe and Co when oxygen vacancies are introduced into STFC, and those changes could actually define many characteristics of the $\delta$-$(x,y)$-stoichiometry.\\ 
%%%%%%%%%%%%%%%%%%%%%%%%%%%%%%%%%%%%%%%%%%%%%%%%%%%%%%%%%%%%%%%%%%%%%%%%%%%%%%%%%%%%%%%%%%%%%%%%%%%%%%%%%%%%
%%%%%%%%%%%%%%%%%%%%%%%%%%%%%%%%%%%%%%%%%%%%%%%%%%%%%%%%%%%%%%%%%%%%%%%%%%%%%%%%%%%%%%%%%%%%%%%%%%%%%%%%%%%%
Figure \ref{figure8} shows that in five out of six configurations  the Co ions are more highly charged than the Fe ions. In all those configurations, one of the vacancies is coordinating the Co, but in the sixth one just the Fe and one Ti are lying in incomplete O-octahedra. Co has a larger electronegativity than Fe, which makes it better able to  capture charge from the unpaired electrons related to the missing oxygens, and the $d$-orbitals enlarged with the occupying charges are more strongly hybridized toward the Co cations.\\

In the sixth configuration at $\sim$188 meV above the lowest-energy state, the same mechanism takes place but with Fe taking the role of Co although, because of the lower Fe electronegativity, the charge difference between Fe and Co is smaller than the previous cases except the ground state to which such difference is comparable. As  seen in Table \ref{table4}, the lowest two defect formation energies are actually for [111]$\delta_{2a}$ and [001]$\delta_{2c}$, which helps to explain the similarity, with [111]$\delta_{2a}$ separated $\sim$188 meV from the $gs$. In Table \ref{table4} $E_f^{\delta}$ = $E_{\text{STFC}}^{\delta}$ - $E_{\text{STFC}}^{\delta=0}$ + $\frac{n_{v}}{2}\mu_{O_2}$, with n$_{v}$ the number of vacancies. Although [111]$\delta_{2a}$ has a slightly lower $E_f^{\delta}$ than [001]$\delta_{2c}$, the formation energies for the distinct Fe-Co arrangements favor [001] over [111] by $\sim$5 times such difference. In the experimental case the deposition conditions are likely to give a variety of combinations of coordination and oxygen vacancy positions, and further analysis would be needed to examine all of the possibilities.\\ 
%%%%%%%%%%%%%%%%%%%%%%%%%%%%%%%%%%%%%%%%%%%%%%%%%%%%%%%%%%%%%%%%%%%%%%%%%%%%%%%%%%%%%%%%%%%%%%%%%%%%%%%%%%%%
\begin{table}[t]
	\captionof{table}{Formation energies $E_f^{\delta}$ of oxygen vacancies with respect to  $\delta=0$ for the configurations in Figure \ref{figure8}.}
	\begin{tabular*}{\linewidth}{c@{\extracolsep{\fill}}c|cc|cc}
		\hline \hline
		\multicolumn{1}{c}{Conf.}   &  $E_f^{\delta}$(eV)& 		\multicolumn{1}{c}{Conf.}   &  $E_f^{\delta}$(eV)& 		\multicolumn{1}{c}{Conf.}   &  $E_f^{\delta}$(eV)\\ 
		\hline 
		\multicolumn{1}{l}{[111] $\delta_{2a}$}  &   10.45 &	\multicolumn{1}{l}{[110] $\delta_{2b}$}  &   10.99 & 	\multicolumn{1}{l}{[001] $\delta_{2b}$}  &   11.36 \\ 
		
		\multicolumn{1}{l}{[111] $\delta_{2c}$}  &   10.51& \multicolumn{1}{l}{[110] $\delta_{2c}$}  &   10.52&	\multicolumn{1}{l}{[001] $\delta_{2c}$}  & 10.49 \\                   
		\hline\hline
	\end{tabular*}
	\label{table4}
\end{table}
%%%%%%%%%%%%%%%%%%%%%%%%%%%%%%%%%%%%%%%%%%%%%%%%%%%%%%%%%%%%%%%%%%%%%%%%%%%%%%%%%%%%%%%%%%%%%%%%%%%%%%%%%%%%

From the magnetism viewpoint the valence state changes that we have discussed here, and which are related to the charge redistribution and structural distortions described above, suggest that the   Fe ions  stabilize  the low spin-state reached by the Co irrespective of the distance $d_{Fe-Co}$, and of the relative locations of the vacancies. The heteropolar bonding described by the hybridization of orbitals require two or more different electronegativities and in this case the difference between the Fe and Co electronic affinity is sufficient for the Fe and Co to  collect to different extents the excess charge generated by the oxygen vacancies. The Co ions present more charge in the 3d orbitals and therefore less total magnetic moment as the low-spin 
state is favored, as corroborated by the correlation between  Figure \ref{figure8} and  Table \ref{table3}. Besides, the interaction between the Fe and Co ions within the covalency framework requires the well known super-exchange interaction through the oxygens, which is also partially influenced by the $4s$ and $4p$ character of Ti when they are connected through Ti-Vo \cite{Mag-STO-delta,OrbitalSymmetry-STO}. 
This suggests that the preferred low spins of Co ions, at least to the extent of the substitutional values and deficiency studied here, might be used to generate large magnetizations/u.c. through the stabilization of high spin states in its covalent transition metal partners. Such mixing of spin-states has been previously studied 
in STO-based systems \cite{STC, OdefSTF,STFexp, roomt1, roomt2}.\\

On the other hand, for $gs$ in Table \ref{table3} the first states in the conduction band at $\sim+0.8$ eV in Figure \ref{figure7} are given by Ti-$t_{2g}$ states. 
There are two types of  orbitals that contribute, those with Ti as the first neighbor with complete octahedral coordination $O_{6}$ (two Ti), and those that have Fe or Co as the first neighbor or that share a vacancy with Fe or Co  (four Ti). 
The first type provides d$_{yz}$ and d$_{xy}$ orbitals to the conduction band while the second type,  except when the Ti shares the vacancy with Co, contributes  d$_{yz}$. We can now consider the occupied part of these $3d$-Ti orbitals. 
The Ti with Ti-V${_O}$-Co coordination has a slightly larger magnetic moment than the other Ti. Both moments are small, but the trend is consistent.
The origin of this difference is a Co-V${_O}$-Ti hybridization through the $d_{z^{2}}$ orbital, which appears to have changed the occupancy from $t_{2g}^{0}e_{g}^{0}$ to $t_{2g}^{1}e_{g}^{0}$; this agrees with the fact that the magnetic moment of Co ions tends to increase when coordinated by oxygen vacancies \cite{STC, OdefSTF}. 
Moreover, this agrees with the change in valence of Co observed in strained oxygen-deficient SrCoO$_3$ \cite{StrainControl}. 
It is an interesting question whether applying strain or generating strain magnetoelastically in a specific direction that is presumed 
to compete with the V${_O}$-induced chemical strain along the Ti-Vo-Co axis could tune the transport properties governed by the $3d$-Ti orbitals.
%%%%%%%%%%%%%%%%%%%%%%%%%%%%%%%%%%%%%%%%%%%%%%%%%%%%%%%%%%%%%%%%%%%%%%%%%%%%%%%%%%%%%%%%%%%%%%%%%%%%%%%%%%%%
%%%%%%%%%%%%%%%%%%%%%%%%%%%%%%%%%%%%%%%%%%%%%%%%%%%%%%%%%%%%%%%%%%%%%%%%%%%%%%%%%%%%%%%%%%%%%%%%%%%%%%%%%%%%
\section{Magnetic, Structural and Band Gap properties versus $\delta$}

\begin{figure}[b]
	\includegraphics[width=1\linewidth]{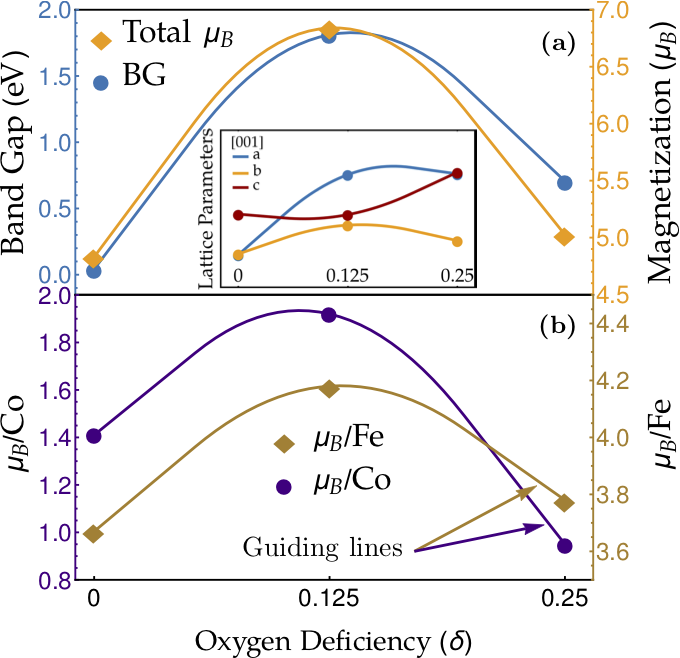}
	\caption{Progression of the band gap, lattice parameter and magnetization for the $gs$ of STFC with $\delta = \{0.0, 0.125, 0.25\}$. }
	\label{figure9}
\end{figure}

Figure \ref{figure9} summarizes the results obtained from our calculations. Figure \ref{figure9} (a) shows that the magnetization as a function of the oxygen 
deficiency increases from $\delta=0$ to $\delta=0.125$ and then decreases for $\delta=0.25$. This qualitatively resembles the results recently reported for STF \cite{OdefSTF}, where the magnetization showed a maximum at a particular base pressure during deposition. In each case the magnetic response of the 
Co and Fe ions to the O-vacancies is fairly similar, both increasingly contributing to the magnetization, as observed in Figure \ref{figure9} (b) for $\delta<0.125$, 
but for larger $\delta$ values the magnetic moment decreases, faster for Co  than for Fe. Figure \ref{figure9} (a) also shows the evolution of the band-gap with 
$\delta$, suggesting that STFC could be turned into a magnetic semiconductor given the correct O-deficiency.\\

From the structural viewpoint, the inset in Figure \ref{figure9} (a) shows the evolution of the lattice parameters for the [001] $gs$ in Tables \ref{table1}, \ref{table2} 
and \ref{table3}. The structures corresponding to those configurations have been depicted in the insets of Figures \ref{figure4} and \ref{figure5}. These last figures illustrate the complex distortions experienced by these perovskites. Moreover, many structures among the ones we have analyzed become non-centrosymmetric due 
to A/B-sites displacements and bending of O$_{5}$ octahedra planes, rather than by just rigid rotation of $O_{6}$ octahedra as usually seen in ABO$_{3}$ $\delta=0$ perovskites. A charge redistribution accompanies these displacements, as discussed in previous sections. This may provide the ingredients necessary 
for the possible appearance of ferroelectric order, as recently suggested in STO-based materials \cite{Brovko:2017hh,OdefSTF}.\\ 
%%%%%%%%%%%%%%%%%%%%%%%%%%%%%%%%%%%%%%%%%%%%%%%%%%%%%%%%%%%%%%%%%%%%%%%%%%%%%%%%%%%%%%%%%%%%%%%%%%%%%%%%%%%%
%%%%%%%%%%%%%%%%%%%%%%%%%%%%%%%%%%%%%%%%%%%%%%%%%%%%%%%%%%%%%%%%%%%%%%%%%%%%%%%%%%%%%%%%%%%%%%%%%%%%%%%%%%%%
The tendencies of the properties in Figure \ref{figure9} are qualitatively robust when we consider charging effects 
due to the vacancies. It is beyond the scope of this article to conduct a full analysis of charged point defects, particularly because  the experimental results 
for STF and STC cited above suggest that cations could stabilize valence states congruent with neutral stoichiometry. However, we analyze a few cases for which the excess or shortage of electrons due to the vacancy are mimicked as charges not localized at the center of the resulting defect but distributed within the cell. Plane-wave-based point-charged-vacancies  along with an analysis of the electric order parameter will be considered elsewhere.\\

%%%%%%%%%%%%%%%%%%%%%%%%%%%%%%%%%%%%%%%%%%%%%%%%%%%%%%%%%%%%%%%%%%%%%%%%%%%%%%%%%%%%%%%%%%%%%%%%%%%%%%%%%%%%
\begin{figure}[t]
	\includegraphics[width=1\linewidth]{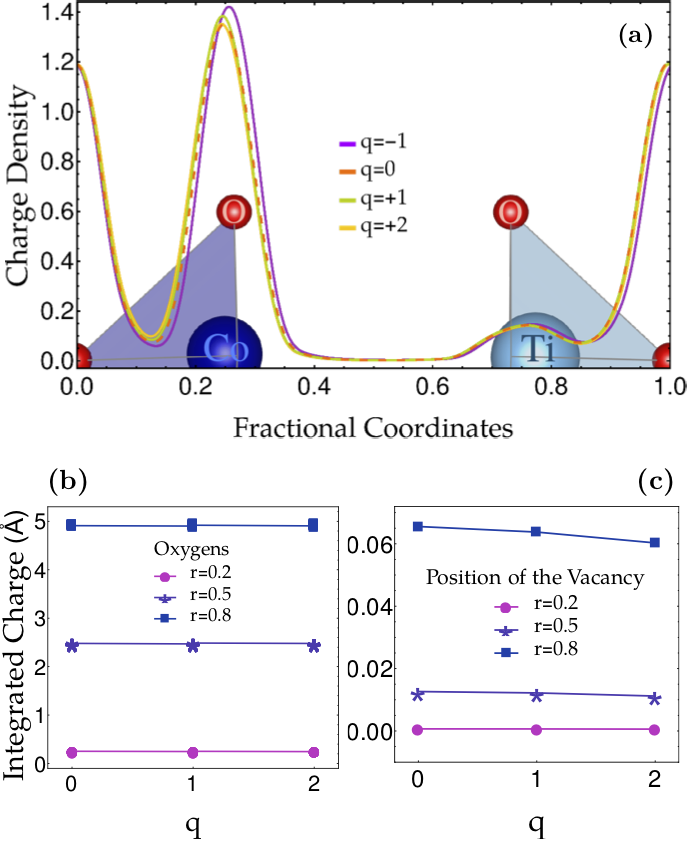}
	\caption{(a) The line charge density containing Co-V$_\text{O}$-Ti  for [001] $\delta_{1b}$ cells with charge q=0,+1 and +2. The corresponding line structure is shown in the background. The integrated charge at different radii can be seen for (b) the position of all oxygens in the cell, and (c) the absent oxygen respectively.}
	\label{figure10}
\end{figure}
%%%%%%%%%%%%%%%%%%%%%%%%%%%%%%%%%%%%%%%%%%%%%%%%%%%%%%%%%%%%%%%%%%%%%%%%%%%%%%%%%%%%%%%%%%%%%%%%%%%%%%%%%%%%
Figure \ref{figure10} (a) shows the charge density for [001]$\delta_{1b}$ evaluated along the Co-$V_{O}$-Ti axis. Three different charges are associated with the O-deficient cells, $q=0,+1,+2$. We can see that at the center of the vacancy between the Co and the Ti ions the stabilized charge has a negligible value compared to what is present at the Co and O locations.  The two oxygens exhibit similar charges. The density charge around the Ti ion is off-centered and slightly shifted toward the vacancy center though the plateau region is $\sim{1.5}$ times wider than the oxygen radii. The lower panels in Figure \ref{figure10} show the integrated charge as observed at certain distances from the center of the oxygens (b) and from the center of the vacancy (c). It is clear that the integrated charge is independent of the initial structural charge and that it increases the same with radius $r$ for all $q$ values. At the vacancy position the integrated charge is at least two orders of magnitude smaller though the increase with the integration radius  mimics the case of the oxygen. Also, the increase of the integrated charge with radius is not approximately linear, as in panel (b), but it has a larger increment from $r=0.5$ to $r=0.8$, which is most likely due to the proximity of neighboring cations that are off-centered causing a similar effect to the one observed in the Ti ion of Figure \ref{figure10} (a).\\

The magnetic properties of a charged one-vacancy supercell were analyzed and summarized in Table \ref{table5}, and the results compared to the ones in Table \ref{table1} and Table \ref{table2}. For $q=+1$, the magnetization/u.c. is $\sim 15\%$ lower than the neutral O-deficient supercell due to the decrease in  the Fe magnetic moment. While the Co ion remains in a $t_{2g}^{5}e_{g}^{2}$ valence spin state, the Fe cation changes from a half-filled $d$ shell to a $t_{2g}^{3}e_{g}^{1}$ valence spin state. In the case of $q=-1$, the magnetization/u.c. is also $\sim 15\%$ lower than the neutral O-deficient supercell but in this case the decrease is due to the change of the Co magnetic moment. While the Fe ion remains in the $t_{2g}^{3}e_{g}^{2}$ valence spin state, the Co cation changes from a $t_{2g}^{5}e_{g}^{2}$ valence state to a $t_{2g}^{6}e_{g}^{1}$. It is clear that charging or discharging the O-deficient supercells does not significantly change  the result in terms of magnetic ordering or saturation magnetization but decreases the magnetization by the magnetic moment of one electron.  Also, adding a charge ($q=-1$) to the system mimics the opposite effect of creating a vacancy in a $\delta=0$ system i.e., it tends to vacate the $e_{g}$ sub-shell. The behavior of STFC is qualitatively the same if we consider $q=+2$; the Fe moment decreases while the Co magnetic moment tends to be stabilized in a Co$^{2+}$. The effect described for the Co and Fe cations is partially a result of their different electronegativities. Once we add an electron to the system Co tends to capture it preferentially compared to Fe. On the contrary, if electrons are to be removed, it tends to be easier to remove those hybridizing the Fe cations.\\
%%%%%%%%%%%%%%%%%%%%%%%%%%%%%%%%%%%%%%%%%%%%%%%%%%%%%%%%%%%%%%%%%%%%%%%%%%%%%%%%%%%%%%%%%%%%%%%%%%%%%%%%%%%%
\begin{table}[h]
	\centering
	\captionof{table}{Magnetism of initially $q$-charged [001] $\delta_{1b}$ unit cell.}
	\begin{tabular}{l@{\extracolsep{\fill}}ccccccc}
		\hline \hline
		\multicolumn{2}{c}{q}   &  $\mu_B$/Fe & $\mu_B$/Co & $\mu_B/u.c.$ & E$_{bg}$(eV)\\ \hline
		\multicolumn{2}{l}{-1}   &    4.17   &   0.98 & 5.83  & 1.62\\
		\multicolumn{2}{l}{ 0}   &    4.18   &   1.92  & 6.84  & 1.81\\
		\multicolumn{2}{l}{+1}&     3.70   &   2.02 & 5.80 & 0.02\\
		%		\multicolumn{2}{l}{+2} & 3.608 & 2.591 & 7  / 6.734 &6.992 & 8.194 & 0&457.9351 \\  
		\hline\hline
	\end{tabular}
	\label{table5}
\end{table}
%%%%%%%%%%%%%%%%%%%%%%%%%%%%%%%%%%%%%%%%%%%%%%%%%%%%%%%%%%%%%%%%%%%%%%%%%%%%%%%%%%%%%%%%%%%%%%%%%%%%%%%%%%%%

In the case of the band-gap, $q=+1$ represents a shrinking of the gap, which is triggered by a new empty sub-orbital in Fe-$e_{g}$ that acts like a spin-filter between $d_{x^{2}-y^{2}}$ and $d_{z^{2}}$ for an Fe-O hybridization. For STFC, the Fe cations provide a path to promote conduction by creating in-gap states, which is a STF characteristic \cite{OdefSTF}. This is different in the $q=-1$ case in which the Ti cations through Co-V$_O$-Ti present one available sub-orbital, previously covalently bonded, that is pushed toward the gap partially due to the effect of the additional electron in the Co d-shell that has a 2p-O character. While these results suggest that the band-gap properties could persist under the introduction or removal of carriers, enough carriers should be able to close the gap if the O-deficiency is not adjusted to balance the metallic effect created by the introduction of charges. \\

Given that charged-defect states associated with oxygen vacancies are not uncommon \cite{superiso_sto,supersemi_sto,superinter_bet_oxi,nonstoi_grain_sto}, we tested the robustness of the system properties against doped-bulk calculations for representative cases. The results seem to corroborate the stability of the magnetic properties though small changes in the Fe and Co magnetic moments are observed, with one saturation magnetization changing by an effective electron spin-moment corresponding to the balancing of the alternating Fe and Co $t_{2g}e_{g}$ population for $+q$ and $-q$ respectively, such a balance  reflecting their different electronegativities. No strong evidence regarding tendencies to charge localization around the vacancy site was detected beside the off-centered charge around the Ti, Fe and Co ions, however, their effects are included in the d-orbital hybridization. This suggests the need for further studies of doping effects as well as the effect of spin-orbit coupling.

%%%%%%%%%%%%%%%%%%%%%%%%%%%%%%%%%%%%%%%%%%%%%%%%%
%%%%%%%%%%%%%%%%%%%%%%%%%%%%%%%%%%%%%%%%%%%%%%%%%
%%%%%%%%%%
%%%%%%%%%%%%%%%%%%%%%%%%%%%%%%%%%%%%%%%%%%%%%%%%%
%%%%%%%%%%%%%%%%%%%%%%%%%%%%%%%%%%%%%%%%%%%%%%%%%
%%%%%%%%%%
\section{Conclusions}
By using hybrid first-principles calculations (HSE), several representative configurations of STFC at different levels of oxygen-deficiency $\delta$ were studied. Electronic properties, stabilized structures, band-gaps and the valence spin-states of the transition metal ions as well as the spin ordering were determined within a DFT $t_{2g}e_{g}$ orbital occupation framework and discussed focusing on the magnetic behavior, which is compared to experimental evidence and theoretical predictions for STFC. \\

Out of all the possible Fe-Co arrangements, the [001] configuration seems to be preferred regardless of the level of oxygen deficiency considered here. A connection between the Fe-Co cation distance and the band gap was found, suggesting that an engineered crystal deposition that could promote tetragonal distortions along the cation axis would promote semiconductor behavior of STFC. \\

We found that for $\delta = 0$, the system is composed of mixed valence states of Fe$^{2+}$ and Co$^{4+}$, in high and low spin states respectively. The accommodation of a single oxygen vacancy in the cell, in coordination with either Fe, Co or both, triggers a charge redistribution that is manifested in a Fe/Co d-suborbital repopulation. This promotes an increase in the local magnetic moments, turning them into Fe$^{3+}$ and Co$^{2+}$ high spins, and therefore effectively raising the saturation magnetization, as well as widening the band-gap, which goes from $50$ meV in the $\delta=0$ $gs$ to $1.8$ eV for the $\delta=0.125$ $gs$. On the other hand, doubling the vacancy concentration of the cell reduces the band-gap and valence states to high spin Fe$^{2+}$ and low spin Co$^{2+}$, diminishing the overall magnetization.\\

The changes of the valence spin state are accompanied by a charge reorganization that resembles the self-regulatory response \cite{chargedensi_vs_oxida_states} seen in other perovskites though we have shown the d-orbital charges change. Also, the band-gap behavior  observed here could be considered to be related to the large energy gap shifting found in perovskite semiconductors \cite{Yan:2013ja}. However, the hybridization of the $t_{2g}e_{g}$-p orbitals in Co-V$_O$-Ti and Co-V$_O$-Fe  dominates the widening or shrinking of the band gap in STFC, usually inserting mid-gap states. \\

Therefore, STFC behaves as a ferromagnetic semiconductor for which the O-deficiency  as well as the Fe-Co arrangement in the unit cell could be used to modulate the magnetism, band gap and structural properties. Among configurations containing the same kinds of vacancies the magnetism and band gap vary within a small range, suggesting that technological developments based on the control of the specific cation distribution and O-vacancies location should be possible. For instance, the larger band gaps found at intermediate $\delta$, together with the B-site shifting, folding of the O$_{5}$ and the charge redistribution observed, are most likely indicators of the appearance of a ferroelectric order parameter in oxygen-deficient STFC.

%+++++++++++++++++++++++++++++++++++++++++++++++++++++++++++++++++++++++++++++++++++++
%+++++++++++++++++++++++++++++++++++++++++++++++++++++++++++++++++++++++++++++++++++++ 

\section{Acknowledgments}
M. A. Opazo acknowledges support from PIIC 004/2017, UTFSM-DGIIP Master Scholarship, and CONICYT Magister Nacional 22182311 2018. 
J. M. Florez acknowledges support from Fondecyt Iniciaci\'on 11130128 and DGIIP-USM. 
C. A. Ross acknowledges support from NSF DMR1419807. 
S.P. Ong acknowledges support from the Materials Project,
funded by the U.S. Department of Energy, Office of Science, Office of Basic Energy Sciences, Materials Sciences and Engineering Division under Contract No. DE-AC02-05-CH11231: Materials Project program KC23MP.
M. A. Opazo and J. M. Florez thank E. A. Cort\'es for his contribution to the discussion of this paper.

%\input{bblfile.bbl}
%\clearpage
%+++++++++++++++++++++++++++++++++++++++++++++++++++++++++++++++++++++++++++++++++++++
%+++++++++++++++++++++++++++++++++++++++++++++++++++++++++++++++++++++++++++++++++++++ 
%+++++++++++++++++++++++++++++++++++++++++++++++++++++++++++++++++++++++++++++++++++++ 
%\bibliographystyle{aipnum4-1.bst}
%\bibliography{biblio_replies.bib}

\end{document}